\documentclass[manuscript]{aastex62}
\usepackage{amsmath}
\graphicspath{{./}{figures/}}

\shorttitle{Repeated Impact-Driven Plume Formation On Enceladus Over Myr Timescales}
\shortauthors{Siraj \& Loeb}

\begin{document}

\title{Repeated Impact-Driven Plume Formation On Enceladus Over Megayear Timescales}

\email{amir.siraj@cfa.harvard.edu, aloeb@cfa.harvard.edu}

\author{Amir Siraj}
\affil{Department of Astronomy, Harvard University, 60 Garden Street, Cambridge, MA 02138, USA}

\author{Abraham Loeb}
\affiliation{Department of Astronomy, Harvard University, 60 Garden Street, Cambridge, MA 02138, USA}




\begin{abstract}
Water plumes erupting from the `tiger stripe' features on the south pole of Enceladus are thought to connect to a global subsurface ocean. Proposed origins for the initial stress necessary to form the `tiger stripes' include a giant impact, which would require true polar wander to explain the location of the plumes if the impact did not occur at the South Pole, or tensile stresses, which would require volumetric expansion associated with partial freezing of the subsurface ocean. In addition to the requirement of a fine-tuned impact, true polar wander, or partial freezing of the subsurface ocean, a further issue with these hypotheses is that the `tiger stripes' may be short-lived. We show here that impact resurfacing can seal off plumes and mass loss can lead to their compression and closure over $\sim 1 \mathrm{\;Myr}$. Since plumes are observed at present, a mechanism by which new plumes can be generated every $\sim 1 \mathrm{\;Myr}$ and by which such plumes are most likely to form at the south pole is needed. We propose and investigate the possibility that impacts constitute an adequate repeating source for the continual instigation of fractures and plumes. We find that the rate of impacts on Enceladus suggests the formation of $\sim 10^3$ independent plume systems per Gyr, the vast majority on the south pole, and is consistent with the Cassini-derived age of the south pole for a lunar-like bombardment history, our estimates of fracture lifetimes, and with the needed parameters for parallel fracture propagation. The model favors a bombardment history similar to that of Triton over one more similar to that of the Galilean satellites.

\end{abstract}



\section{Introduction}
Enceladus' physical libration suggests the existence of a global subsurface ocean \citep{2009Icar..203..541H, 2016Icar..264...37T}. The South Pole Terrain (SPT) of Enceladus contains unique `tiger stripe' features emitting plumes of water, thought to be fractures in the moon's icy shell connecting to a subsurface ocean \citep{2011GeoRL..3818201P, 2014AJ....148...45P, 2016Icar..272..319I, 2016PNAS..113.3972K, 2017NatAs...1E..63L, 2019Icar..332..111H}. The plumes have made Enceladus a target for astrobiological research and a candidate in the search for a second genesis \citep{2015Natur.519..207H, 2017LPI....48.1401P, 2017Sci...356..155W, 2017AsBio..17..840B, 2017AsBio..17..852J, 2018eims.book..437M, 2020AsBio..20..179K, 2020GeoRL..4785885G, 2020SSRv..216....9T}.

The lack of cratering on the SPT indicates a young age relative to the rest of the surface of Enceladus, $\sim 1 \mathrm{\; Myr}$ for a lunar-like impact chronology \citep{2006Sci...311.1393P}. A lunar-like impact chronology could potentially indicate the strength of the Nice model \citep{2007AJ....134.1790M, 2011AJ....142..152L, 2012ApJ...744L...3B, 2012AJ....144..117N}.

The giant impact hypothesis generally requires true polar wander to explain the location of the SPT \citep{2006Natur.441..614N, 2017LPI....48.1955R, 2017Icar..295...46T}, which may invoke fine-tuning arguments for the formation of a single hotspot, as observed \citep{2019arXiv191212554K}. Specifically, reorientation is needed to move the present-day south polar terrain to the polar location regardless of where it formed. However, if it formed in place, true polar wander would not be required, although an explanation would be necessary to explain why formation occurred specifically at the south pole. The \cite{2019NatAs.tmp....4H} model presents a plausible formation scenario for the tiger stripes as a result of ridge loading and bending stresses after one of the central fractures is initiated, however the tensile stresses \citep{2007GeoRL..34.7202M, 2009Icar..199..536R} necessary to initiate the first fracture would require a few hundred meters of ocean freezing \citep{2019NatAs.tmp....4H}. While it is not clear whether the thermal history of Enceladus would support such a freezing \citep{2011AGUFM.P11B1596C, 2014P&SS..104..185C, 2015Icar..250...32T}, this paper presents two arguments for why plumes may be short-lived, one based on mass loss, and the other due to impact resurfacing, and thus by the Copernican principle, there should be a time-independent and repeatable origin for plumes.

A repeatable source for an initial fracture giving rise to plumes is cratering by heliocentric debris. Here, we explore the cratering rate and the implication for the rate at which plumes on different parts of Enceladus could be formed. We also show that plumes likely have limited lifetimes due to resurfacing caused directly by cratering from smaller objects as well as compression of the icy shell due to mass loss.

Our discussion is structured as follows. In Section \ref{sec:impactfracture}, we explore the mechanism for fracture formation from impactors and evaluate the predicted rates of fracture for different pole thicknesses of ice on Enceladus. In Section \ref{sec:lifetime} we discuss the lifetime of fractures, which is limited by compression due to global mass loss and by the rate of impact resurfacing. Finally, in Section \ref{sec:disc}, we summarize key implications of our model.

\section{Impact Fracturing}
\label{sec:impactfracture}
\subsection{Fracture formation}
\label{sec:fractureformation}

An icy impactor with mass $M$, radius $R$, and speed $v$ penetrates an icy shell with thickness $L$ to a depth, $\sim 4R/3$, where it has roughly encountered an equal volume of material over approximately the crossing time of that distance, $\sim 4R/3v$. The impactor releases a large part of the lost kinetic energy as a pressure wave that emanates from the submerged surface and propagates throughout the icy shell, decaying as $(x/R)^{-1.8}$, where x is the distance from the impactor's surface \citep{2008JGRE..11311002S}, although there is considerable uncertainty in the decay exponent \citep{1997Icar..127..408P}. The pressure at the bottom of the icy shell is,

\begin{equation}
\label{frac}
    P_d \sim \frac{3 M v^2 }{8 \pi R^{1.2} L^{1.8}} \; \; .
\end{equation}
Adopting the dynamic shear strength of water ice to be $\sim 27 \mathrm{\; MPa}$ \citep{2000GeoRL..27..305A}, we consider an icy shell fractured by an impact when $P_d \gtrsim 27 \mathrm{\; MPa}$. While the transient crater may be shallow relative to the total thickness of the ice shell, the pressure wave with a magnitude larger than the dynamic shear strength of water ice should result in fracturing at more significant depths.

The icy shell of Enceladus is thought to be at least $\sim 20 - 25 \mathrm{\; km}$ \citep{2014AGUFM.P53A3998T, 2019Icar..332..111H} and up to $\sim 40 \mathrm{\; km}$ thick globally \citep{2014Sci...344...78I, 2016DPS....4821401L, 2019AGUFM.P53D3479T}, thinning to $\sim 15 \mathrm{\; km}$ at the north pole and $\sim 9 \mathrm{\; km}$ at the south pole \citep{2019NatAs.tmp....4H}. Adopting these estimates, as well as a fiducial impact velocity of $20 \mathrm{\; km \; s^{-1}}$, an icy impactor with radius $R \gtrsim 65 \mathrm{\; m}$ is capable of fracturing the south pole, one with $R \gtrsim 110 \mathrm{\; m}$ could fracture the north pole, and one with $R \gtrsim 280 \mathrm{\; m}$ is able to fracture the icy shell of any other region on Enceladus, assuming a global thickness of $\sim 40 \mathrm{\; km}$. Faster impact speeds would allow for smaller impactors. Given the tradeoff between impact size and velocity, there are a range of impacts that deliver equivalent energy to the subsurface. However, the distribution of the energy is not necessarily identical for impacts delivering the same amount of energy.

The assumption of a normal trajectory is justified as follows. The infinitesimal element of solid angle is $d[\sin \theta]$, and since the average value of $\sin \theta$ over the range of possible impact angles is $(2/\pi)$, the typical factor of the ice shell thickness that is experienced by an impactor due to the angle of approach is ($2/\pi)^{-1} = (\pi/2)$. We use equation \eqref{frac} to translate this into the correction in impactor radius relative to the impactor radius necessary to fracture the ice shell on a normal trajectory: $((\pi/2)^{1.8})^{1/3} \sim 1.3$, so only an order-unity (30\%) increase in size relative to a normal trajectory when we average over all impact angles. Averaging over all possible impact angles does not change significantly the size-frequency distribution under the assumption of normal incidence.

\subsection{Fracture rate}
\label{sec:fracturerate}

\begin{figure}
  \centering
  \includegraphics[width=0.97\linewidth]{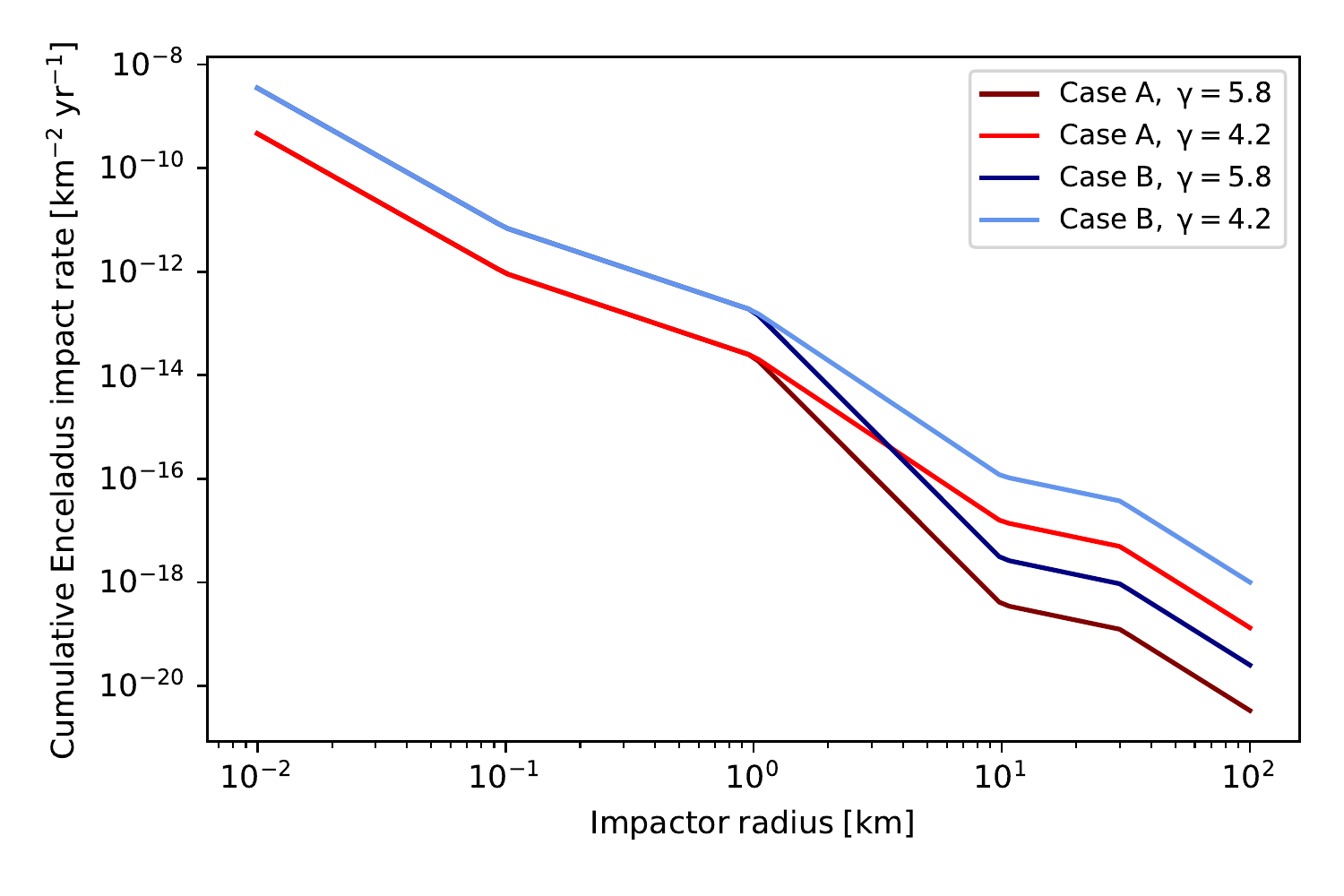}
    \caption{Cumulative Enceladus impact rate as a function of impactor radius.}
    \label{fig:sizedist}
\end{figure}

The current cratering rate of Enceladus by heliocentric debris with radius $R > 0.75 \mathrm{\; km}$ is $3.7 \times 10^{-14} \mathrm{\; km^{-2} \; yr^{-1}}$ for Case A\footnote{Consistent with the distribution of small comets at Jupiter and its satellites, and requiring planetocentric debris at Saturn.} and $2.8 \times 10^{-13} \mathrm{\; km^{-2} \; yr^{-1}}$ for Case B\footnote{Consistent with the distribution of small craters on Triton, and not necessarily requiring planetocentric debris.}, each regarded as uncertain to a factor of 4 \citep{2009sfch.book..613D}. We adopt the following size distribution for impactors, as constrained independently from Kuiper Belt collisional evolution models and from the crater size distributions of Saturnian moons: $N(>R) \propto R^{1-q}$, where

\begin{equation}
\label{size}
    q \sim
  \begin{cases}
    4.0 , & \text{$R > 30 \mathrm{\;km},$} \\
    2.0 , & \text{$10 \mathrm{\;km} < R < 30 \mathrm{\;km},$} \\
    \gamma , & \text{$1 \mathrm{\;km} < R < 10 \mathrm{\;km},$} \\
    2.6 , & \text{$0.1 \mathrm{\; km} < R < 1 \mathrm{\;km},$} \\
    3.7 , & \text{$ R < 0.1 \mathrm{\;km},$} \\
  \end{cases}
\end{equation}
and where we consider the cases of $\gamma = 5.8$ \citep{2012LPICo1667.6348M} and of $\gamma = 4.2$ \citep{2013AJ....146...36S}. We note that there exists uncertainty of the extrapolation below the quoted lower limit of $R \sim 10 \mathrm{\; m}$ \citep{2012LPICo1667.6348M, 2013AJ....146...36S}. This size distribution is adopted for both Case A and Case B. Figure \ref{fig:sizedist} indicates the cumulative Enceladus impact rate as a function of imapctor radius.

\begin{figure}
  \centering
  \includegraphics[width=0.97\linewidth]{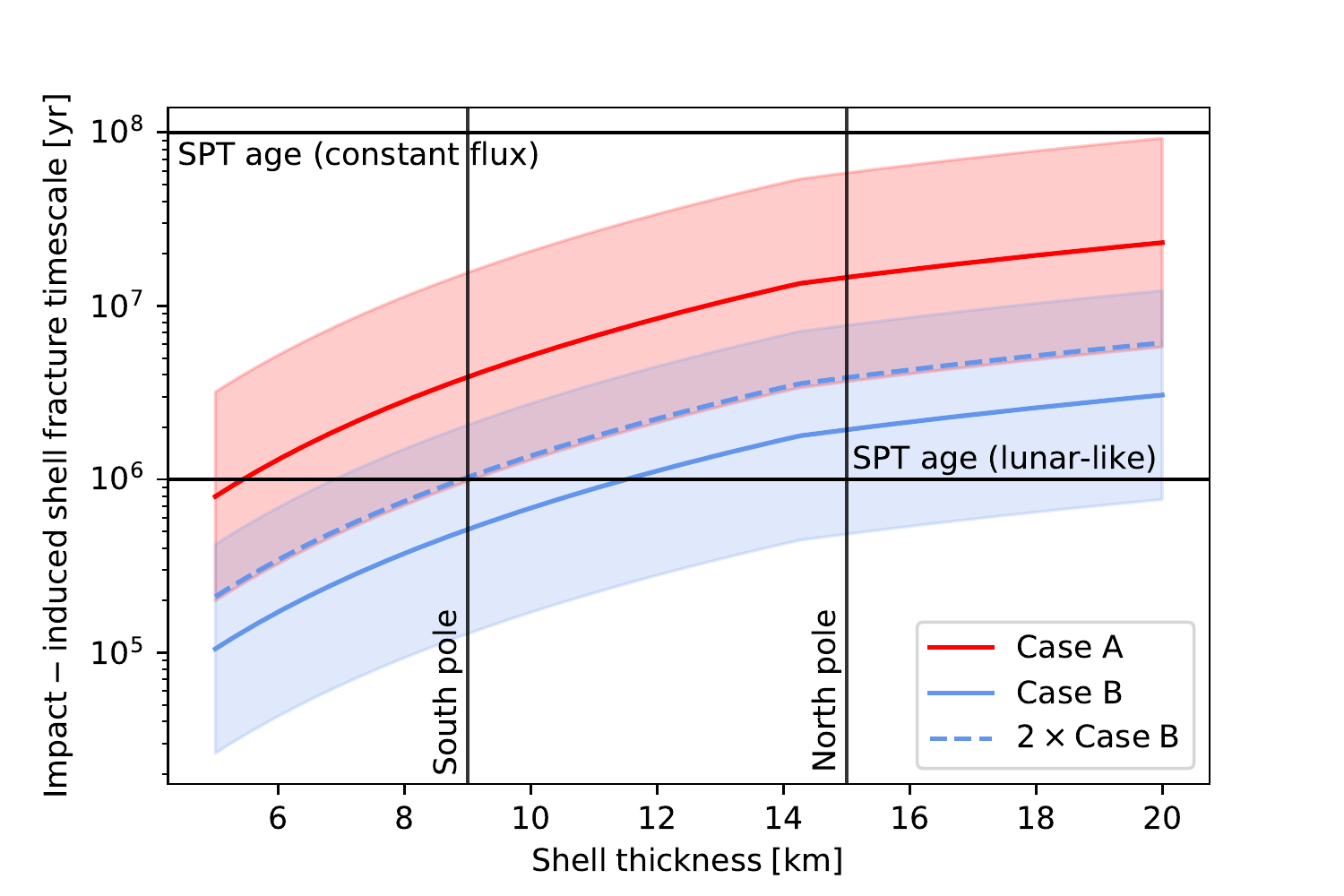}
    \caption{Timescale of impact-induced shell fracture as a function of shell thickness for a pole (constituting $\sim 0.1$ of the total surface area) on Enceladus. The remainder of Enceladus' surface (covered by a $\sim 40 \; \mathrm{km}$ shell) would fracture on a timescale of $\sim 3 \times 10^6 \; \mathrm{yr}$ for $2 \times$ Case B (the best fit for the Cassini-measured age of the SPT).}
    \label{fig:plot1}
\end{figure}

We use the cumulative size distribution of impactors, combined with our fracture model, to compute the fracture timescale as a function of shell thickness for a pole on Enceladus, as shown in Figure \ref{fig:plot1}. Specifically, we derive the necessary impactor size that would result in a pressure $\gtrsim 27 \mathrm{\; MPa}$ at the base of the ice shell with thickness $L$ (see equation \eqref{frac}), and we subsequently compute the associated impact rate for both Case A and Case B, applying the size distribution described in equation \eqref{size}. Only the `lunar-like' SPT age of $\sim 1 \mathrm{\; Myr}$ derived by Cassini is compatible with the impactor rates, and itself is more compatible with Case B than Case A, with a best-fit of `$2 \times$ Case B'. $2 \times$ Case B implies a fracture timescale of $\sim 10^6 \mathrm{\; yr}$ for the south pole, $\sim 5 \times 10^6 \mathrm{\; yr}$ for the north pole, and $\sim 3 \times 10^6 \mathrm{\; yr}$ for the remainder of Enceladus, since the surface area of non-polar regions greatly exceeds that of polar regions by definition. The fracture timescale is defined as the timescale to fracture the ice of a given thickness and given area. These results suggest that over 1 Gyr, $\sim 10^3$ fractures form on Enceladus, with the majority occurring on the South Pole, and an enhanced fraction ($\gtrsim 10 \times$) relative to the rest of Enceladus occuring on the North Pole. While tiger stripes are expected to form everywhere on Encladus, they should form more often at the South Pole than anywhere else. Once a single fracture is formed, it has been demonstrated that parallel fractures can be formed on short ($\lesssim 1 \; \mathrm{Myr}$) timescales \citep{2019NatAs.tmp....4H}.

\section{Fracture Lifetime}
\label{sec:lifetime}

\subsection{Fracture healing due to mass loss}

\begin{figure}
  \centering
  \includegraphics[width=0.97\linewidth]{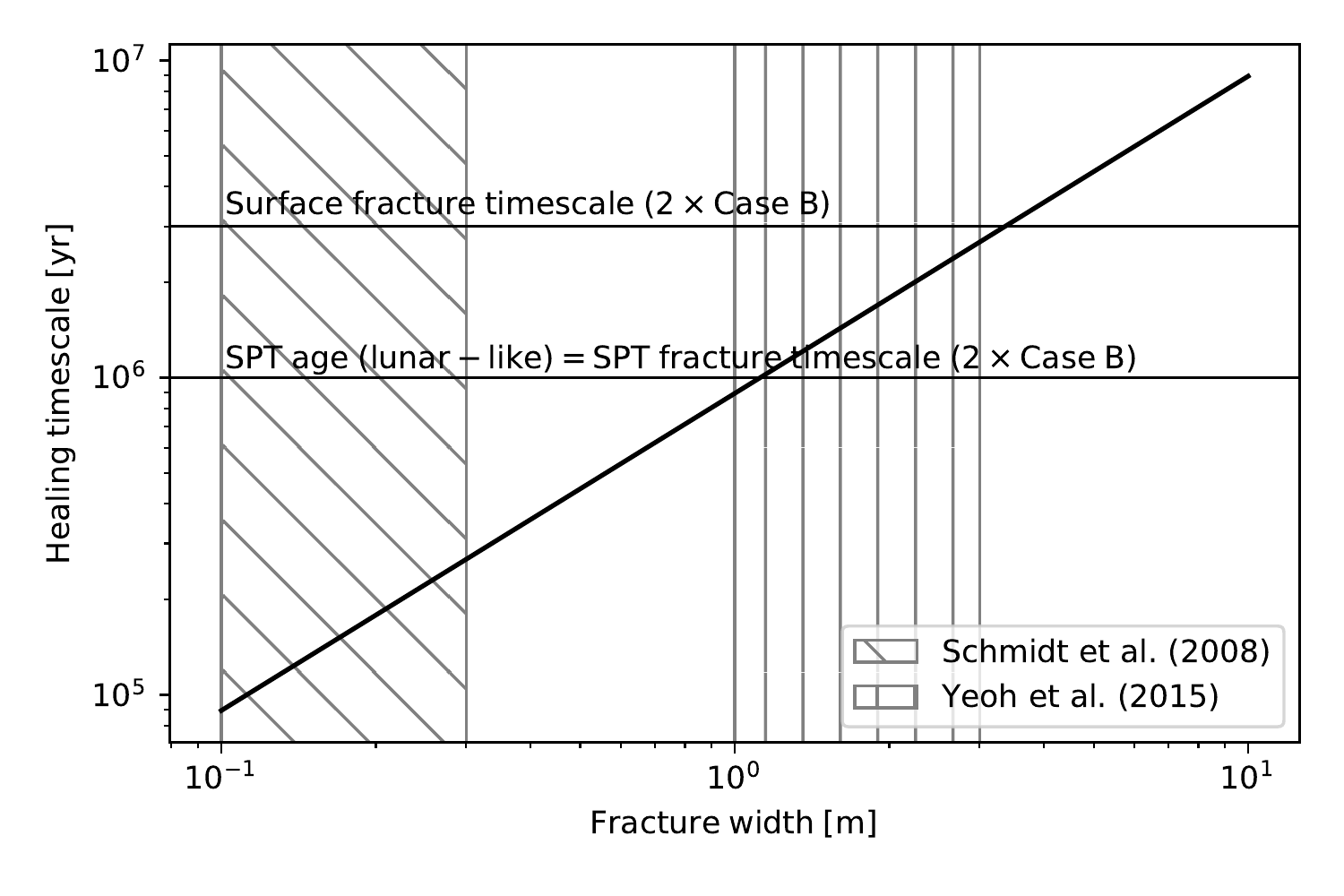}
    \caption{Healing timescale (in yr) as a function of fracture width (in m) due to mass loss from fractures, taken to be $\sim 200 \; \mathrm{kg \; s^{-1}}$. The distance scale on which the resulting compression can heal a fracture of a given width is taken to be the distance between the `tiger stripes', $\sim 35 \mathrm{km}$.}
    \label{fig:plot4}
\end{figure}

The Enceladus plumes are estimated to eject $\mathrm{H_2 O}$ at a rate of, $\dot{M} \sim 200 \mathrm{\; kg \; s^{-1}}$ \citep{2011GeoRL..3811202H}, although the historical rate is otherwise unconstrained and the propagation of the current rate over the lifetime of the solar system would result in significant mass loss on $\sim Gyr$ timescales. If the historical mass loss rate was lower than the present rate by a given factor, the rate of fracture healing due to mass loss would be reduced by that same factor, potentially allowing for plume systems that had been resurfaced but not closed throughout the entire thickness of the Encledaus shell. The outflow flux per tiger stripe is,

\begin{equation}
\dot{M} l /(4 \pi R_e^2) \; \; , 
\end{equation}
and this must equal,

\begin{equation}
\rho_w (dA/dt) \; \; , 
\end{equation}
where $dA/dt = 2 \pi R_e \dot{w}$, $\dot{w}$ is the healing rate, $\rho_w$ is the density of water, and $R_E$ is the radius of Enceladus. Given that the `tiger stripes' are each separated by the distance, $l \sim 35 \mathrm{\; km}$, lateral compression of each fracture due to mass loss occurs at a rate of,

\begin{equation}
\label{flux}
    \dot{w} \sim \frac{\dot{M} l}{\rho_w (4\pi R_E^2) (2\pi R_E)} \; \;.
\end{equation}
The ice shell will compress along with the shrinking ocean, since the pressure at the bottom of the ice shell due to gravity, $L \rho_w g_E \sim \mathrm{a \; few \; MPa}$, is greater than the tensile failure limit \citep{2019NatAs.tmp....4H}, $\sim 0.1 - 1 \mathrm{\; MPa}$, of the ice shell. The resulting healing rate is $\sim 1 \mathrm{m \; Myr^{-1}}$, signifying that the compression due to mass loss is capable of healing a $\sim 1 \mathrm{\; m}$ fracture (plume nozzle size predicted by \citealt{2015Icar..253..205Y}) in the SPT over $\sim 1 \mathrm{\; Myr}$. The approximate healing timescale as a function of fracture width is indicated in Figure \ref{fig:plot4}. The nozzles located within the `tiger stripes' from which mass is ejected have been estimated to have widths of $\sim 0.1 \mathrm{\; m}$ or of $\sim 1 \mathrm{\; m}$ \citep{2008Natur.451..685S, 2015Icar..253..205Y}, and could remain open by a variety of processes \citep{2016Icar..272..319I, 2018eims.book..163S}. Nozzles with widths $\sim 1 \mathrm{\; m}$ \citep{2015Icar..253..205Y} would be completely consistent with the fact that the plumes from the south pole are still observed today and that they are currently unique on Enceladus' surface. Compression due to mass loss is based upon mass conservation, since the pressure equilibration timescale within the global ocean is very short relative to the compression timescale. Therefore, it is a global effect, and does not depend on the local ocean depth. We note that partial freezing of the ocean could prevent global compression due to mass loss on short timescales. 

\subsection{Impact Resurfacing}
We propose that small impactors, which melt surface ice upon impact, could seal off these nozzles and arrest the lifetimes of any resulting plumes. While the surface pressure of the vents is $\sim 1 \mathrm{\; kPa}$, thin sheets of ice fail at pressures of $\gtrsim 1 \mathrm{\; MPa}$ \citep{Zhang2012, WU2015155, QI2017}, making them effective at sealing off nozzles once formed.

For a $\sim 20 \mathrm{\; km \; s^{-1}}$ impactor into $\sim 50 \mathrm{\; K}$ ice, the total melt mass is $\sim 10^2$ times that of the impactor \citep{2011Icar..214..724K}. Following the scaling relations for $\mathrm{H_2 O}$ ice \citep{2011Icar..214..724K}, we find that the concentration of impact melt in a small plug in the crater floor \citep{2011Icar..214...67S} has surface area,

\begin{equation}
    A \sim 1.3 \times 10^{4} \mathrm{\; m^2} \; \left(\frac{R}{10 \mathrm{\; m}} \right)^{7/4} \left(\frac{v}{20 \mathrm{\; km \; s^{-1}}} \right)^{2/3} \; \;,
\end{equation}
and thickness,

\begin{equation}
    \tau \sim 1.4 \mathrm{\; m} \; \left(\frac{R}{10 \mathrm{\; m}} \right)^{7/4} \left(\frac{v}{20 \mathrm{\; km \; s^{-1}}} \right)^{2/3} \; \;.
\end{equation}

\begin{figure}
  \centering
  \includegraphics[width=0.97\linewidth]{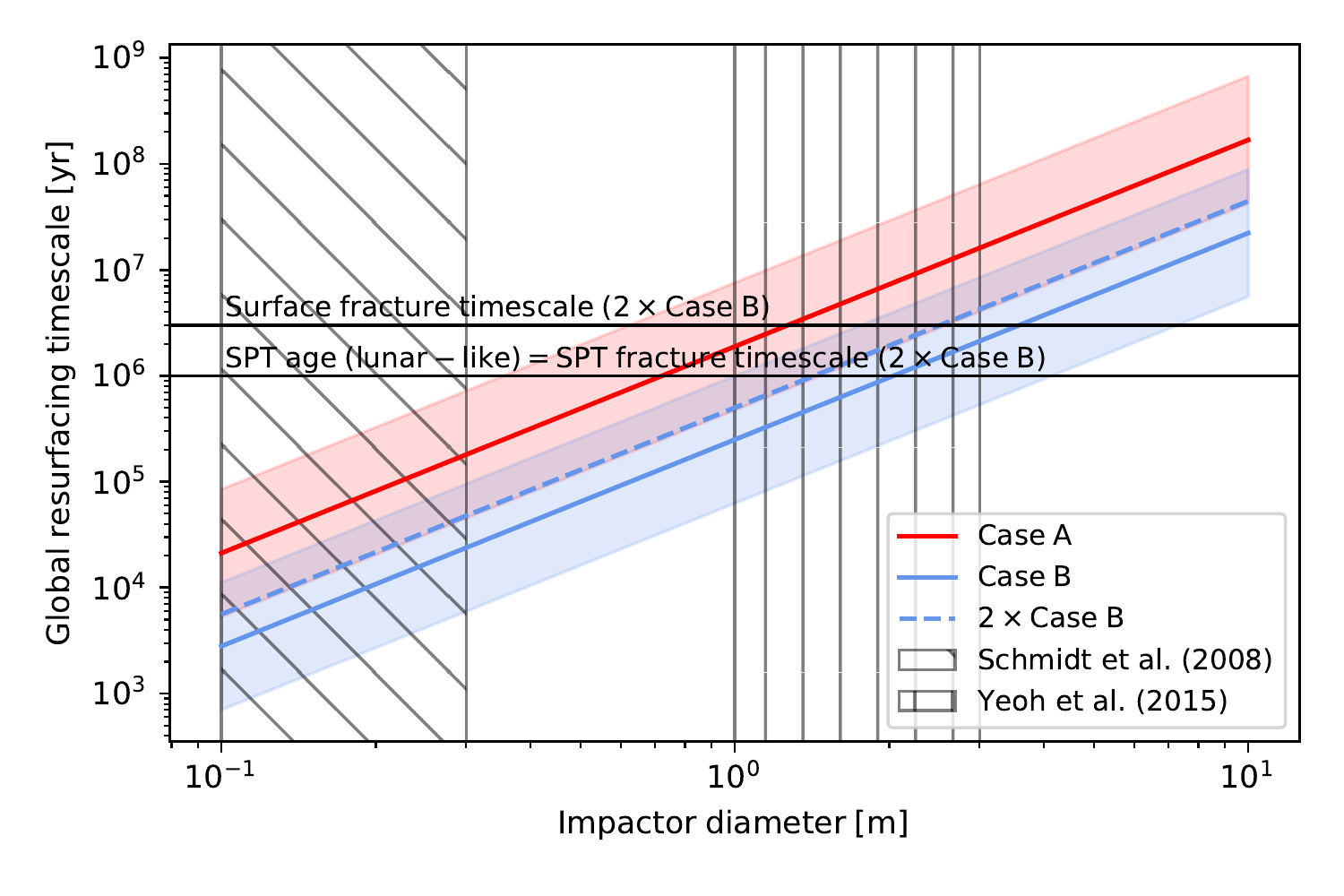}
    \caption{Global resurfacing timescale as a function of impactor diameter.}
    \label{fig:plot2}
\end{figure}

The timescales on which these melt plugs cover the entire surface of Enceladus as a function of impactor diameter considered (using the differential size distribution of impactors extrapolated to the relevant size) are shown for Case A and Case B in Figure \ref{fig:plot2}, with the SPT age and the $2 \times$ Case B surface fracture timescale for reference. The width of the nozzle corresponds directly to the impactor size scale considered, since smaller impactors are more numerous but ones that are smaller than the width of the nozzle could fall inside of it. We would not expect to observe crater morphology from those impactors on the timescale of a resurfacing since our requirement that melt plugs cover the entire surface dictates that the larger craters must overlap with several others, eroding the topography of such craters and any pre-existing topography with elevations smaller than the total resurfacing depth, across the entire surface of Enceladus. Compatible timescales are longer than the SPT age, otherwise we would not currently observe plumes, and shorter than the Case B surface fracture timescale, which would ensure that any trace of old fractures on the surface would be obscured by an ice sheet. These conditions are consistent with both Case A and Case B. Melt produced by the impacts that cause fracturing, since confined to the surface, could also seal off any pre-existing features in its immediate environment.

\begin{figure}
  \centering
  \includegraphics[width=0.97\linewidth]{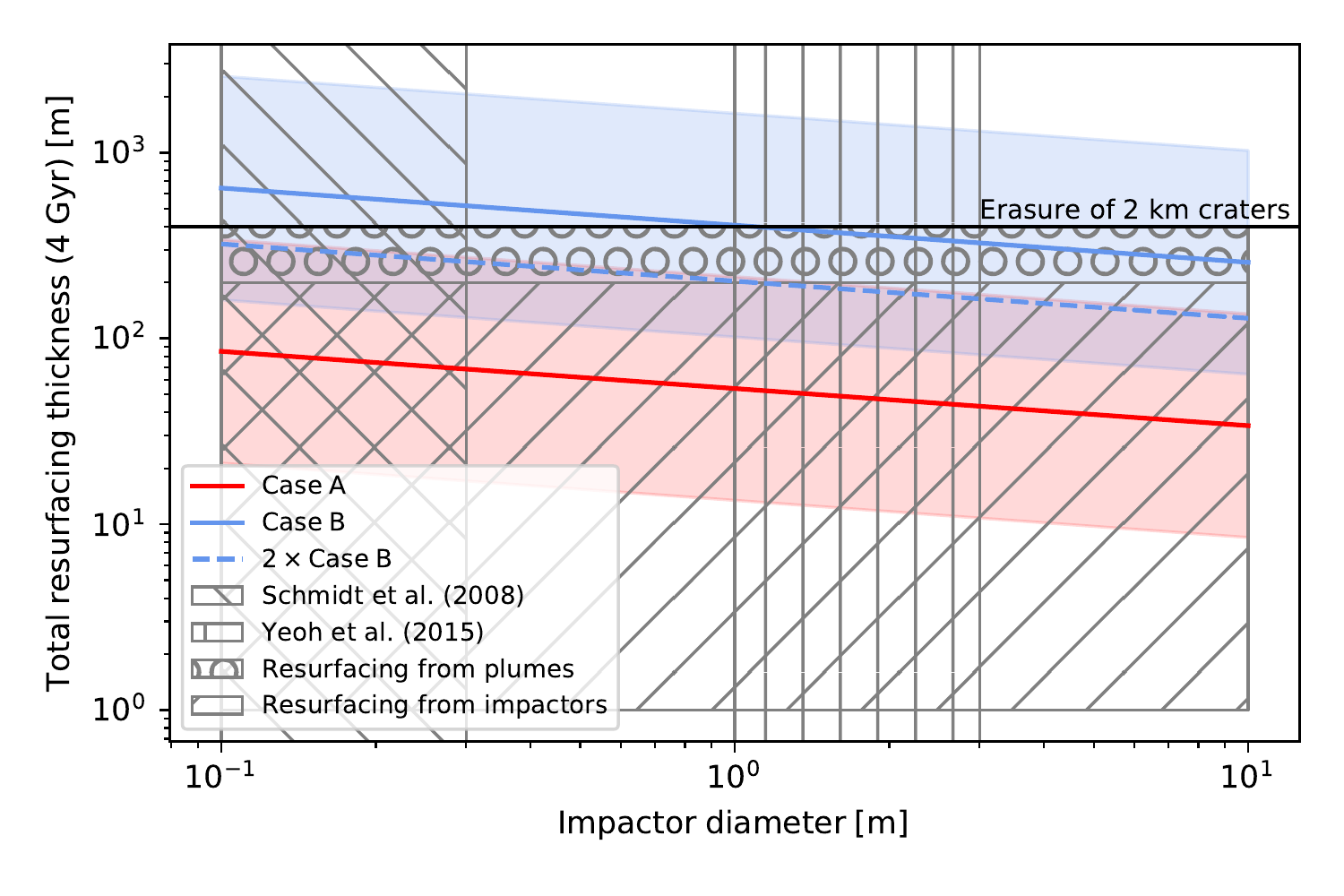}
    \caption{Total resurfacing thickness from impacts over 4 Gyr as a function of impactor diameter.}
    \label{fig:plot3}
\end{figure}

We also compute the total implied thickness of ice due to resurfacing over 4 Gyr. Enceladus has a deficiency of craters with diameters $\lesssim 2 \mathrm{\;km}$ and $\gtrsim 6 \mathrm{\;km}$ compared to the other satellites of Saturn. While the low abundance of large craters could be attributed to viscous relaxation, the $\gtrsim 2 \mathrm{\;km}$ craters are thought to be filled in by SPT plume material \citep{2009Icar..202..656K}, which could contribute $\sim 2 \times 10^2 - \sim 4 \times 10^2 \mathrm{\; m}$ of material over 4 Gyr \citep{2007Icar..188..154T}. We use the crater depth-diameter relationship as observed on Galilean icy satellites to estimate the depth of a $D \sim 2 \mathrm{\; km}$ crater to be $\sim 400 \mathrm{\; m}$ \citep{2002Natur.417..419S}. In order to match the observed dearth of $\gtrsim 2 \mathrm{\;km}$ craters, impactor resurfacing could have contributed $\lesssim 2 \times 10^2 \mathrm{\; m}$ of material, as indicated in Figure \ref{fig:plot3}. Both Case A and Case B are consistent with this condition.

\section{Discussion}
\label{sec:disc}

Our model of repeatedly generated plumes on Enceladus predicts $\sim 10^3$ independent plume systems per Gyr, the vast majority originating from the South pole with the rest originating elsewhere on the surface, including an enhanced occurence density at the north pole due to its relatively thin ice. Our fracturing model is dependent on ice thickness, however the healing timescales for fractures in thin and thick ice are similar, since both compression due to mass loss and resurfacing due to small impactors are global, rather than local, phenomena. As models of the current state as well as the evolution of Enceladus' icy shell improve, our model's predictions will become more precise. Additionally, the particular mechanism by which plumes operate once a fracture has been initiated is not the focus of this work.

Our model is consistent with the age of the SPT derived from Cassini data for a lunar-like bombardment history, and inconsistent with a constant flux, making the existence of an early bombardment an important test for our model \citep{2006Sci...311.1393P}. As a result, Case B is favored over Case A, implying that the bombardment history on Enceladus is more similar to that of Triton than to that of the Galilean satellites \citep{2009sfch.book..613D}. Our model favors nozzle widths of order $1 \mathrm{\; m}$ \citep{2015Icar..253..205Y} over nozzle widths of order $0.1 \mathrm{\; m}$ \citep{2008Natur.451..685S}.

Without fine-tuning, this model for impact-driven plume formation over Myr timescales is consistent with all observations, indicating its strength. In particular, this model does not require the fine-tuning that is required by many models that require that the plume on Enceladus to be formed within the past few Myr ($\mathrm{\sim 0.1 \%}$ of the lifetime of the solar system). The sensitivity of our model to the true abundance and size distribution of the small-body population near Saturn for impact resurfacing will serve to assess its validity. Specifically, future constraints on the size distribution of small impactors on Enceladus will inform the model parameters and overall model plausibility. Additionally, geological evidence of dead plumes on Enceladus would strengthen our model. But already now, our model is compatible with the wide range of surface ages and features observed on Enceladus \citep{2006Sci...311.1393P, 2011GeoRL..3818201P}, as well as the predictions made by \cite{2019NatAs.tmp....4H} relating to the younger ages of the non-central fractures in the SPT, in addition to the \cite{2019arXiv191212554K} symmetry-breaking model, and the encouraging recent observations of possible evidence of geological resurfacing near the north pole \citep{2020Icar..34913848R}. The magnitude of heat flux from Enceladus can be explained by tidal dissipation \citep{2016MNRAS.458.3867F}, and thus does not require a fracture-related explanation.

\section*{Acknowledgements}
This work was supported in part by a grant from the Breakthrough Prize Foundation. 





\bibliography{bib}{}
\bibliographystyle{aasjournal}



\end{document}